\documentclass[conference,a4paper]{IEEEtran}
\IEEEoverridecommandlockouts
\usepackage{cite}
\usepackage{amsmath,amssymb,amsfonts}
\usepackage{algorithmic}
\usepackage{graphicx}
\usepackage{textcomp}
\usepackage{multirow}
\usepackage{booktabs}
\usepackage{color}
\usepackage[T1]{fontenc}
\usepackage[table,xcdraw]{xcolor}
\def\BibTeX{{\rm B\kern-.05em{\sc i\kern-.025em b}\kern-.08em
T\kern-.1667em\lower.7ex\hbox{E}\kern-.125emX}}
\begin{document}

\title{LSTN: A Linear Model of Industrial Production Process for Demand Response
}

\author{\IEEEauthorblockN{Ruike Lyu$^{1}$, Hongye Guo$^{1}$, Yuanjie Zheng$^{2}$, Yunlong Bai$^{3}$, and Qixin Chen$^{1}$}
\IEEEauthorblockA{Department of Electrical Engineering, Tsinghua University, Beijing, China$^{1}$}
\IEEEauthorblockA{State Grid Corporation of China, Beijing, China$^{2}$}
\IEEEauthorblockA{State Grid Anhui Electric Power Co., Ltd., Hefei, China$^{3}$}
\IEEEauthorblockA{Email: qxchen@tsinghua.edu.cn$^{1}$}

\thanks{This work was supported by the Science and Technology Program of State Grid Corporation of China under Grant 5108-202218280A-2-378-XG. 979-8-3503-9678-2/23/\$31.00 ©2023 IEEE}

}

\maketitle

\begin{abstract}
Industrial production modeling provides operational constraints for industrial users participating in demand response (DR) programs. Conventional modeling of the production process introduces binary variables to model the discrete operating points of industrial equipment, which can be computationally infeasible in large-scale DR applications. To reasonably model industrial users' operational constraints while balancing computational complexity and modeling accuracy, we developed a linear model of the industrial production process for evaluating DR applications. Numerical results verify the accuracy of the proposed model and its great improvement in computational efficiency over competing approaches.
\end{abstract}


\section{Introduction}

It has been widely recognized that demand response (DR) has huge potential for securing electricity in the power system. Exploiting the flexibility of existing demand-side resources can reduce the enormous cost of investment in renewable energy capacities, flexible power generation resources, and grid expansion~\cite{zhuo_cost_2022}. Among demand side resources, the proportion of energy consumption by industrial users is much higher than that of other load types, which results in greater potential for power adjustment by industrial consumers\cite{samad_smart_2012}. Moreover, the energy consumption characteristics of industrial users are more regular, so they may be the first to achieve large-scale DR and provide greater flexibility for the power grid~\cite{wohlfarth_demand_2020}.

To avoid economic losses, the technical constraints of industrial production processes and the production targets need to be included in the energy consumption strategy for industrial-user DR.
Ref.~\cite{ding_demand_2014} proposed a DR energy management scheme for industrial loads based on the state-task network (STN) model, which determined the scheduling of processing tasks according to day-ahead hourly electricity prices.
Ref.~\cite{zhang_industrial_2015} focused on the steel plant and optimized its scheduling, maximizing the profit in both the energy and the spinning reserve markets.
Ref.~\cite{yu_real-time_2016} applied robust optimization to address the uncertainty of future prices, with a model that made real-time industrial load scheduling decisions while considering future load obligations.
In general, an interesting trend of industrial production process modeling for DR is to use general-purpose models such as STN~\cite{lu_multi-agent_2020} and resource-task network (RTN) models built on STN~\cite{castro_resourcetask_2013}.

In these commonly used STN and RTN models, since the operation points of industrial equipment are usually discretely categorized into “on” and “off” states, binary variables are inevitably introduced to model the production process of a factory. As a result, modeling DR often becomes a mixed integer linear program (MILP) that essentially requires NP-hard combinatorial optimization. Thus, growing complexity drastically increases computation time. However, with the increase in computing power and the improvement of general accelerating methods for MILP, a DR strategy for a single factory based on STN (and including necessary simplification) can be solved in a few seconds~\cite{lu_data-driven_2021}. Nevertheless, we believe that linear modeling of the industrial production process is still necessary for the following reasons:
\paragraph*{\textbf{To overcome computation complexity issues}}
In the future, a large number of factories may be aggregated to participate in DR, such as through virtual power plants, and the solution complexity of MILP increases exponentially with the problem size. In other words, such joint energy consumption optimization can be computationally infeasible. Thus, existing production process modeling methods will hinder future large-scale industrial DR, and a more computationally efficient model needs to be developed.
\paragraph*{\textbf{To increase production model accuracy}}
Although the operation points of industrial equipment are discrete, most devices can be switched rapidly~\cite{zhang_demand_2018}. Thus, the energy consumption of these devices over a piratical period of time can essentially be considered as continuous. Therefore, the commonly used discrete time steps of an hour are also suboptimal. Especially in DR scenarios where the focus is on the hourly energy consumption of industrial users (rather than the minute-level energy consumption), continuous-time models are likely to be more advantageous~\cite{nolde_electrical_2010}.
\paragraph*{\textbf{To be easier to embed into existing models}}
On the one hand, as mentioned above, if the existing production process models based on binary variables are directly embedded into hierarchical models such as economic dispatch and market clearing algorithms, the complexity of the upper model increases significantly. On the other hand, the optimality conditions of discrete models can be mathematically intractable, which makes conventional pricing design methodologies based on a mathematical program with equilibrium constraints (MPEC) ~\cite{cortez_demand_2023} more challenging. In short, a modeling method that can be adapted into the existing power system management paradigm is needed.

The main concept of the existing linear or continuous modeling of the industrial production process is to model the time to switch the operation point of industrial equipment as a continuous variable instead of equal-length time intervals~\cite{floudas_continuous-time_2004}. However, this modeling is inconsistent with the implementation of DR programs in the context of deregulated electricity markets, where electricity prices are based on scheduled time intervals, e.g., hourly day-ahead prices.

Here, we develop a linear STN (LSTN) for industrial production process modeling oriented to DR applications. Starting from the production process modeling, a linear model of a single industrial user is formulated including the desired amount of final product, the production, consumption, and storage of materials, and the energy consumption characteristics of each production task.
In the context of DR, the discrete variables for the operating points of production tasks are replaced by their operating times, which are modeled as continuous variables to establish a linear model. Then, we present the application of LSTN for industrial user DR, establishing an aggregation-decomposition framework that is practical for evaluating DR scenarios with a large number of participating factories while accommodating privacy concerns.

The remainder of this paper is organized as follows: Section~\ref{sec_LSTN} describes the modeling of LSTN. Section~\ref{sec_application} presents the application of LSTN in industrial user DR. Section~\ref{sec_numerical} presents the numerical tests with realistic data. Section~\ref{sec_conclusion} contains the conclusions and prospects for future research.

\section{Industrial Production Process Modeling}\label{sec_LSTN}

In this section, we present the production process modeling of industrial users based on the STN model and the linearization of the original model. Compared with STN, the main contribution of LSTN is to focus on the energy consumption characteristics of the production process in an industrial user. In other words, it focuses on the temporal-coupling constraints of an industrial user's energy consumption, rather than the specific operation of the devices.

\begin{figure}[!t]
  \centering
  \includegraphics[width=2.5in]{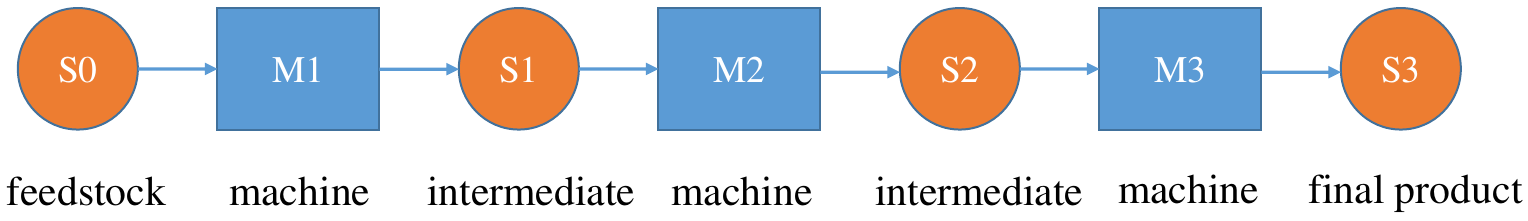}
\caption{A common state-task network model for industrial production process. ${\rm M}_i$ and ${\rm S_i}$ represent the $i$th machine (“task”') and its product (“state”), both included explicitly as network nodes. Here, the task can be an actual industrial device or several devices for the same production objective. The states can represent the feedstock, intermediate, or final products.}
  \label{fig_LSTN}
\end{figure}

The general-purpose STN (Fig.~\ref{fig_LSTN}) can model a wide range of production processes arising in multiproduct/multipurpose industrial facilities~\cite{kondili_general_1993}.
Consider an arbitrary factory $f \in F$ (for brevity, the subscript $f$ is omitted in this section, e.g., $E_{t}$ for $E_{ft}$). Let $i/i^{\rm end}$ denote the index/total number of production tasks. Let $I^{\rm P}/I^{\rm S}$ denote the set of production tasks/states ($I^{\rm P} = \{1, 2, ..., i^{\rm end}\}$, $I^{\rm S} = \{0\} \cup I^{\rm P}$, $i=0$ for the feedstock). Let $t$ and $T$ denote the index of time intervals and the whole time horizon, respectively ($T = \{0, 1, 2, ..., t^{\rm end}\}$, $t=0$ for the initial or current time interval). Let $k/K_i$ denote the index/set of the operating points of task $i$.

The conventional STN model assumes that industrial devices can only operate at one point within a time interval. In the context of DR, although industrial tasks (equipment) usually operate at discrete points, the time to switch their operating points (e.g., seconds to minutes~\cite{zhang_demand_2018}) is negligible compared to the time interval during which the electricity price is implemented (e.g., an hour). Considering this feature, we treat the operating time of the tasks in the points as a continuous variable to obtain a linear-formed model as follows:

Factory $f$ aims to minimize $Cost$, its total energy cost across $T$:
\begin{equation}\label{primal_cost}
  Cost = \underset{t \in T}{\Sigma}{Pr_{t}} E_{t},
\end{equation}
where $Pr_{t}$ (\$/kWh) is the electricity price and $E_{it}$ (kWh) is the energy consumption of factory $f$ at $t$. $E_{t}$ is decomposed to the energy consumption of the operating points of the tasks:
\begin{equation}\label{primal_constraint_E}
  E_{t} = \underset{i \in I^{\rm P}}{\Sigma} \underset{k \in K_{i}}{\Sigma} P_{ik} \Delta t_{tik}, \ t \in T,
\end{equation}
where $P_{ik}$ (kW) is the electricity consumption power of task $i$ at operating point $k$ and $\Delta t_{tik}$ (h) is the time duration (continuous) that task $i$ is operating at point $k$ within $t$. Naturally, the operation time of a task in all its operating points within a time interval, including the off state, is nonnegative (\ref{primal_constraint_timeNonNegative}) and sums up to the length of the interval $\Delta t$ (\ref{primal_constraint_timeSum}):
\begin{equation}\label{primal_constraint_timeNonNegative}
  \Delta t_{tik} \ge 0, \  t \in T, i \in I^{\rm P}, k \in K_i.
\end{equation}
\begin{equation}\label{primal_constraint_timeSum}
  \underset{k \in K_{i}}{\Sigma} \Delta t_{tik} = \Delta t, \  t \in T, i \in I^{\rm P}.
\end{equation}
Note that we only focus on the duration or the length of time that the devices operate at their different operating points (e.g., “on”/“off”) in the time periods, rather than the specific time to switch operating points. After determining the operating duration, the feasibility of the specific time to switch operating points is guaranteed by (\ref{primal_constraint_timeNonNegative})-(\ref{primal_constraint_timeSum}), and operations are left for the factory to implement. 

Let $S_{ti}$ (kg) denote the amount of material $i$ at the end of time interval $t$. Let $S^{\rm 0/max/tar}_i$ denote the initial/upper limit/target amount of material $i$.
The facility must meet its production target (\ref{primal_constraint_tar}) including a buffer limit (\ref{primal_constraint_storageLimit}), and the initial states of the buffers are given by (\ref{primal_constraint_S0}):
\begin{equation}\label{primal_constraint_tar}
  S_{ti} \ge S^{0}_{i} + S^{\rm tar}_i, \  t = t^{\rm end}, i \in I^{\rm S}.
\end{equation}
\begin{equation}\label{primal_constraint_storageLimit}
  0 \le S_{ti} \le S^{\rm max}_{i}, \  t \in T, i \in I^{\rm S}.
\end{equation}
\begin{equation}\label{primal_constraint_S0}
  S_{ti} = S^{\rm 0}_i, \ i \in I^{\rm S}, t = 0
\end{equation}

Let $G_{ik}/C_{ik}$ (kg/h) denote the material production/consumption rate of task $i$ at operating point $k$. The change in buffer states across time for the feedstock, intermediate, and final products is given by (\ref{primal_constraint_changeofS1}), (\ref{primal_constraint_changeofS2}), and (\ref{primal_constraint_changeofS3}), respectively:
\begin{equation}\label{primal_constraint_changeofS1}
  S_{ti} = S_{(t - 1)i}
  - \underset{k \in K_{i+1}}{\Sigma} C_{(i+1)k}  \Delta t_{t(i+1)k}, \ t \in T, i = 0.
\end{equation}
\begin{equation}\label{primal_constraint_changeofS2}
  \begin{aligned}
    & S_{ti} = S_{(t - 1)i} + \underset{k \in K_{i}}{\Sigma} G_{ik}  \Delta t_{tik}
  - \underset{k \in K_{i+1}}{\Sigma} C_{(i+1)k}  \Delta t_{t(i+1)k}, \\
  & t \in T, i \in I^{\rm P} \setminus \{i^{\rm end}\}.
\end{aligned}
\end{equation}
\begin{equation}\label{primal_constraint_changeofS3}
  S_{ti} = S_{(t - 1)i} + \underset{k \in K_{i}}{\Sigma} G_{ik}  \Delta t_{tik}, \ t \in T, i = i^{\rm end}.
\end{equation}

We instantiate the above model in linear programming (LP) to form the LSTN model, in which the decision variables are $\{\Delta t_{tik} | \forall i, \forall t, \forall k\}$, and the dependent variables are $\{E_{ti}, S_{ti} | \forall i, \forall t\}$.
The problem parameters include electricity prices $\{Pr_t|\forall t\}$ and equipment parameters $\{G_{ik}, C_{ik}, P_{ik}, S^{\rm 0/max/tar}_i | \forall i, \forall k\}$, which can be obtained directly through the device nameplate or the factory inventory.

Although LSTN is derived for devices with discrete operating points, it can also model the operation of nonadjustable devices as well as devices with inherently continuous operating points. We verify that the former can be modeled by only one operating point and the latter can be modeled by two operating points, corresponding to “on” (i.e., operating at rated power) and “off” states.
For devices whose time to switch the operating point is not negligible, binary variables can be introduced in conventional ways. As an alternative, the continuous-time assumption of LSTN can still be used, which requires sacrificing model accuracy for computational efficiency. We will analyze the effect of this compromise in the case study.

\section{Application of LSTN in Industrial User DR}\label{sec_application}

In this part, we present the application method of LSTN in industrial user DR. LSTN models the energy cost and production process constraints of a factory as linear, so the most straightforward application is to directly embed LSTN in existing models with DR. An example is to embed the primal problem of LSTN into an economic dispatch. Another example is to embed the optimality conditions of LSTN into DR pricing problems, such as KKT conditions, dual constraints, and strong duality conditions, which are easy to express and computationally feasible for LP problems. Due to space limitations, we omit describing our implementation of the direct embedding method.

However, we do note that this direct embedding implies a central control architecture that can obtain the internal parameters of the factories, which may not be practical due to privacy concerns.
To further reduce the computational burden and protect the privacy of industrial users, we propose a three-layer control design as follows.

\subsection{Aggregation and Control}

\begin{figure}[!t]
  \centering
  \includegraphics[width=3in]{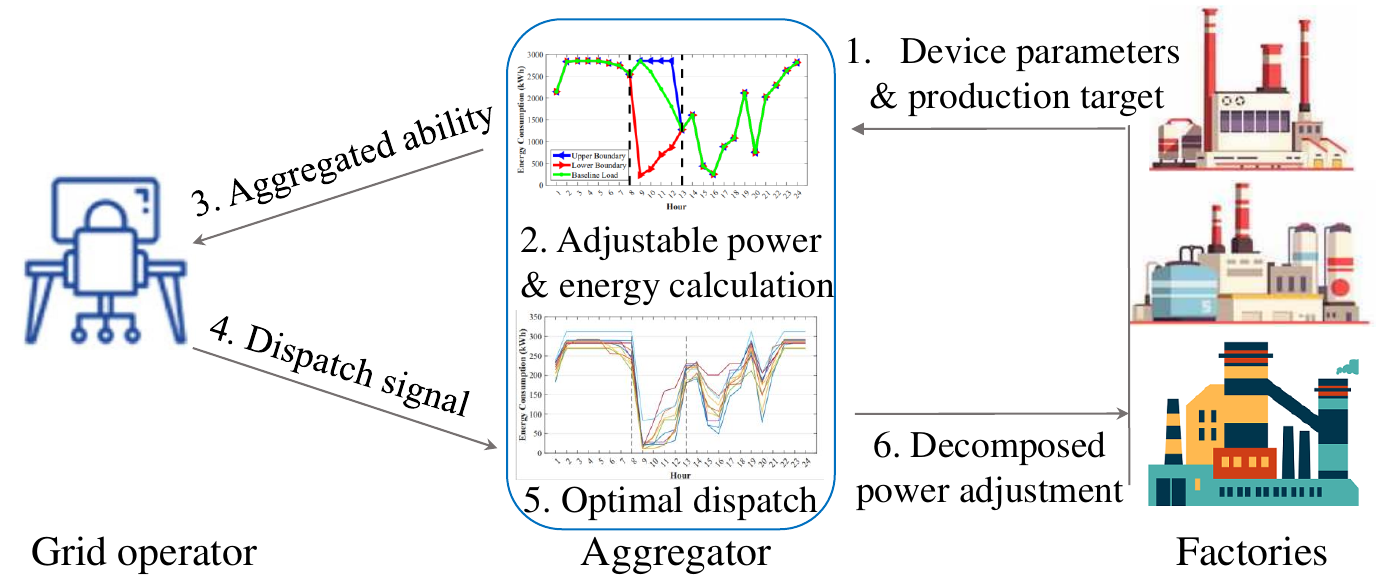}
\caption{Overview of the proposed framework for industrial user DR.}
  \label{fig_overview}
\end{figure}

The aggregator resides between the factories and the grid operator (or electricity market) and calculates the aggregated DR capacity using the parameters of the set of factories $F$, reports this capacity to the grid operator, receives the dispatching signal from the grid, and assigns control signals to the individual factories (Fig.~\ref{fig_overview}).

Different models can be used to describe the DR capacity of factories. Here, we choose the virtual battery (VB) model for illustration. The problem of identifying the optimal VB parameters is to approximate the external characteristics of the aggregation resources with a battery model~\cite{tan_optimal_2023}. The optimization of the model parameters is beyond the scope of this article. Here, we present an intuitive approach for determining the solution.
As the four parameters of the VB, let $\overline{P}^{\rm B}_t/\underline{P}^{\rm B}_t$ and $\overline{E}^{\rm B}_t/\underline{E}^{\rm B}_t$ denote the charging/discharging power capacity and charging/discharging energy capacity at time interval $t \in T^{\rm DR}$, respectively. Without loss of generality, these parameters may have different values at different time intervals, determined by:
\begin{equation}\label{VB_objective}
  {\rm max.} \ \eta^{\top}[\overline{P}^{\rm B}_t;\underline{P}^{\rm B}_t;
  \overline{E}^{\rm B}_t;\underline{E}^{\rm B}_t]
\end{equation}
\begin{equation}\label{VB_cons_individual}
  {\rm s.t.} \ (\ref{primal_constraint_E})-(\ref{primal_constraint_changeofS3}), \ \forall f \in F,
\end{equation}
\begin{equation}\label{VB_cons_P}
    \overline{P}^{\rm B}_{t} \le \underset{f \in F}{\Sigma}(E_{ft} - \tilde{E}_{ft})/\Delta t \le - \underline{P}^{\rm B}_t, 
\end{equation}
\begin{equation}\label{VB_cons_E}
    \overline{E}^{\rm B}_t \le \underset{(\tau \in T^{\rm DR}, \tau \le t)}{\Sigma} \underset{f \in F}{\Sigma}(E_{f\tau} - \tilde{E}_{f\tau}) \le -\underline{E}^{\rm B}_t,
\end{equation}
where $\eta$ is the coefficient vector for keeping the problem compact, e.g., $\eta = [1;0;0;0]$ for calculating $\overline{P}^{\rm B}_{t}$, and the four parameters need to be calculated; $\tilde{E}_{ft}$ is the baseline load of factory $f$ at $t$, e.g., determined by the production scheduling strategy formulated by the factory in the face of day-ahead electricity prices; $T^{\rm DR}$ is the time horizon of DR events, e.g., $T^{\rm DR}=\{1, 2, 3, 4\}$ with the common DR duration of 4 hours~\cite{li_precision_2022}. $T^{\rm DR}$ can be different from $T$ in (\ref{primal_constraint_E})-(\ref{primal_constraint_changeofS3}).
(\ref{VB_cons_individual}) states the energy consumption constraints described by the LSTN for each factory.
It is easy to verify that (\ref{VB_cons_P}) and (\ref{VB_cons_E}) are the definitions of the VB parameters, e.g., substituting $\eta = [1;0;0;0]$ to verify the definition of $\overline{P}^{\rm B}_{t}$.

\subsection{Optimal Dispatch}

The grid operator sends control signals to the aggregator through predefined strategies according to the aggregated DR capacity of the factories as reported via VB parameters. For the grid to determine the control signals, various existing strategies with the participation of conventional batteries in
unit commitment~\cite{wen_frequency_2016}, output smoothing of wind generation~\cite{piphitpattanaprapt_optimal_2015}, economic optimization in the electricity market, and other commonly used models can be directly applied. Therefore, we omit the process of grid-computing the signals and only present the optimal dispatching strategy of the aggregator after receiving the signals.

Upon receiving the control signals $\delta_t, t \in T^{\rm DR}$ (kW), the aggregator solves the following problem to optimize the energy consumption of the factories while responding to the control signals:
\begin{equation}\label{OD_objective}
  {\rm min.} \ \underset{t \in T}{\Sigma} \underset{f \in F}{\Sigma} Cost_{f} + M\underset{t \in T^{\rm DR}}{\Sigma} \Delta E_{t}^{2}
\end{equation}
\begin{equation}\label{OD_cons_individual}
  {\rm s.t.} \ (\ref{primal_cost})-(\ref{primal_constraint_changeofS3}), \ \forall f \in F
\end{equation}
\begin{equation}\label{OD_cons_mismatch}
  \underset{f \in F}{\Sigma} (E_{ft} - \tilde{E}_{ft}) - \delta_{t}\Delta t = \Delta E_t, \ t \in T^{\rm DR}
\end{equation}
where $M$ is a large number, $\Delta E_{t}$ is the energy mismatch defined by (\ref{OD_cons_mismatch}), and the second term in the objective function (\ref{OD_objective}) is a penalty term enforcing that the factories follow the load reduction control signal. Without loss of generality, the control signals within $T^{\rm DR}$ must be followed, while the costs and operating constraints of the factories within the whole time interval $T$ must be considered.
The above aggregating and dispatching models are LPs and can be efficiently solved.

\section{Numerical Results}\label{sec_numerical}

We used Gurobi (V10.0.0) and MATLAB (R2021a) with YALMIP~\cite{Lofberg2004} to solve the optimization problems. Computation was executed on a workstation with an Intel Core i9-10900X CPU (3.7 GHz) and 128 GB RAM.

\subsection{Modeling Accuracy}

\begin{table}[!t]
  \caption{Parameters of the original steel powder manufactory.}
  \label{tab_parameter}
  \centering 
  \begin{tabular}{lllll}
  \toprule
  Name &
  \begin{tabular}[c]{@{}l@{}}Operating\\point\end{tabular} &
  \begin{tabular}[c]{@{}l@{}}Production\\rate\\(Ton/h)\end{tabular} &
  \begin{tabular}[c]{@{}l@{}}Energy\\demand\\(kWh)\end{tabular} &
  \begin{tabular}[c]{@{}l@{}}Buffer\\capacity\\(Ton)\end{tabular} \\ \midrule
                             & off  & 0  & 0  &                          \\ \cline{2-4}
\multirow{-2}{*}{Reduction}  & on & 15 & 75 & \multirow{-2}{*}{100}    \\ \hline
                             & off  & 0  & 0  &                          \\ \cline{2-4}
\multirow{-2}{*}{Atomizer}   & on & 30 & 60 & \multirow{-2}{*}{180}    \\ \hline
                             & off  & 0  & 0  &                          \\ \cline{2-4}
\multirow{-2}{*}{Dehydrator} & on & 15 & 10 & \multirow{-2}{*}{100}    \\ \hline
\rowcolor[HTML]{FFFFFF} 
\cellcolor[HTML]{FFFFFF}     & off  & 0  & 0  & \cellcolor[HTML]{FFFFFF} \\ \cline{2-4}
\multirow{-2}{*}{\cellcolor[HTML]{FFFFFF}Dryer} &
  on &
  15 &
  30 &
  \multirow{-2}{*}{\cellcolor[HTML]{FFFFFF}150} \\ \hline
                             & off  & 0  & 0  &                          \\ \cline{2-4}
\multirow{-2}{*}{Separator}  & on & 15 & 10 & \multirow{-2}{*}{100}    \\ \hline
                             & 1   & 0  & 0  &                          \\ \cline{2-4}
                             & 2   & 10 & 15 &                          \\ \cline{2-4}
\multirow{-3}{*}{Crusher}    & 3   & 15 & 20 & \multirow{-3}{*}{100}    \\ \hline
                             & 1   & 0  & 0  &                          \\ \cline{2-4}
                             & 2   & 10 & 15 &                          \\ \cline{2-4}
\multirow{-3}{*}{Classifier} & 3   & 10 & 25 & \multirow{-3}{*}{150}    \\ \hline
                             & 1   & 0  & 0  &                          \\ \cline{2-4}
                             & 2   & 10 & 6  &                          \\ \cline{2-4}
\multirow{-3}{*}{Blender}    & 3   & 15 & 10 & \multirow{-3}{*}{200} \\  \bottomrule
\end{tabular}
\end{table}

\begin{figure}[!t]
  \centering
  \includegraphics[width=2.5in]{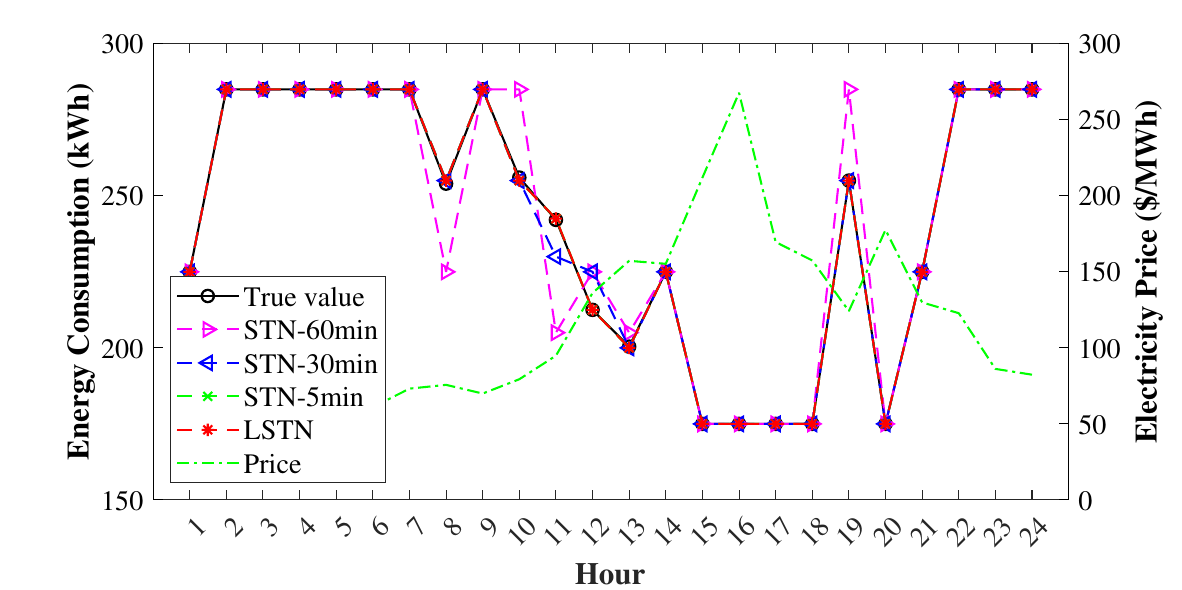}
\caption{Load profile on August 5 given by the compared models.}
  \label{fig_baseline_load}
\end{figure}

We tested the modeling accuracy for the scenario of a single industrial user using hourly electricity price data for August 2022 in the PJM network and the site parameters of a steel powder manufacturing facility ~\cite{lu_data-driven_2021} (Tab.~\ref{tab_parameter}). The production target of the factory is set to 24 times the lowest production rate across the production tasks (i.e., 15 Ton/h), and the initial material is half the buffer limit. We assume that the time required for switching the operating state of the plant equipment is 2 minutes. Therefore, the STN model with a scheduling time interval of 2 minutes (STN-2 min) is treated as an accurate industrial user model, with load profiles and daily energy costs generated using the MILP-based production process model~\cite{lu_data-driven_2021} used as the true values for calculating the error of other models. The errors measured by RMSE in the 31 days of the month are listed in Tab.~\ref{tab_rmse_models}. As an example, Fig.~\ref{fig_baseline_load} presents the load profiles given by each model on August 5.
$${\rm RMSE}=\sqrt{\frac{\sum^N_{i=1} (y^*_i - y_i)^2}{N}}$$

\begin{table}[!t]
  \caption{The performance of the compared models.}
  \label{tab_rmse_models}
  \centering
  \begin{tabular}{ccc}
    \toprule
    Models &
    \begin{tabular}[c]{@{}c@{}}RMSE for load profile (kW)\\ /Ratio to max. value \end{tabular} &
    \begin{tabular}[c]{@{}c@{}}RMSE for energy cost (\$)\\ /Ratio to max. value \end{tabular} \\ \midrule
      STN-60min  & 15.3(5.37\%) & 1.979(0.175\%) \\
  STN-30min & 3.43(1.21\%) & 0.398(0.035\%) \\
  STN-5min  & 0.59(0.21\%) & 0.034(0.003\%) \\
  \textbf{LSTN}  & \textbf{0.30(0.17\%)}   & \textbf{0.033(0.003\%)}  \\ \bottomrule
  \end{tabular}
  \end{table}

Tab.~\ref{tab_rmse_models} shows that the energy consumption error of LSTN is only 0.59 (0.21\% of the maximum load) of the true value (STN-2 min), which is less than the error of 15.3 (5.37\% of the maximum load) of the STN-60 min model. This indicates that assuming the operating time of the industrial devices to be continuous is reasonable, and the derived LSTN model is more accurate than the MILP-based STN model while offering a longer scheduling time interval.

\subsection{Computation Efficiency}

\begin{figure}[!t]
  \centering
  \includegraphics[width=2.5in]{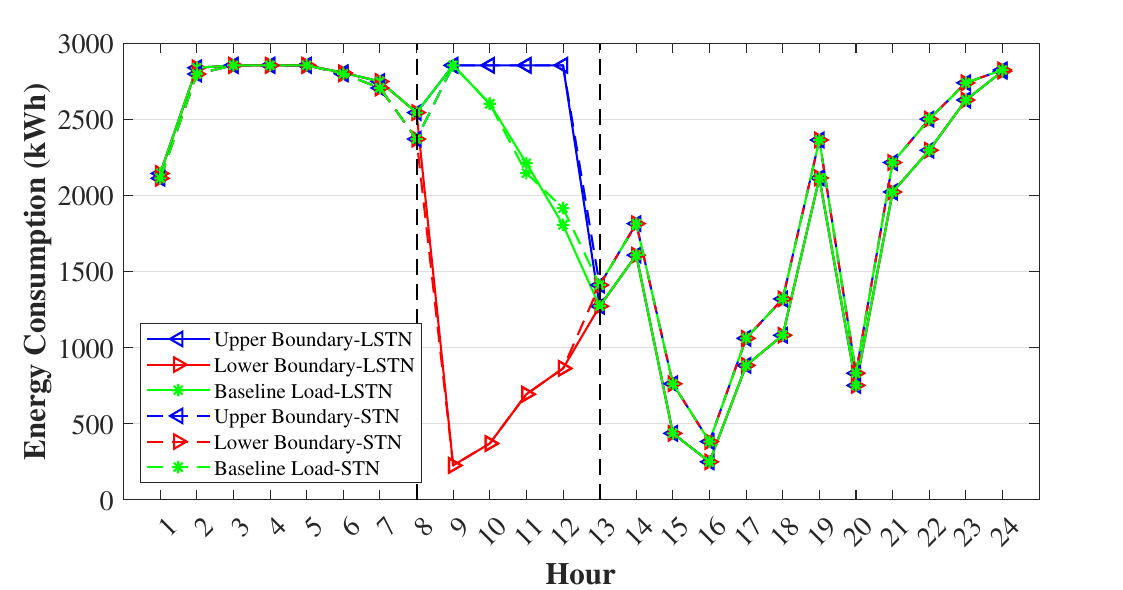}
\caption{Adjustable Boundary of the Aggregated Factories on August 5.}
  \label{fig_boundary}
\end{figure}

To verify the improvement of computational efficiency due to the LSTN approach, we used it to evaluate the adjustable capacity of the aggregated factories and dispatch grid instructions to individual industrial users. The default number of industrial users is 10, and the device parameters are uniformly distributed between [0.8, 1.2] times the values given in Tab.~\ref{tab_parameter}. The DR duration is 4 hours. STN-60 min was also evaluated for comparison because this is the model commonly used and is the most computationally efficient among the MILP-based models. We used the load profile on August 5 as the load baseline and then calculated the adjustable ability of the aggregated factories (Fig.~\ref{fig_boundary}), given by STN-60 min and LSTN. Although the load baselines given by the two models are different, their results for the adjustable boundary are very similar.

\begin{figure}[!t]
  \centering
    \includegraphics[width=1.5in]{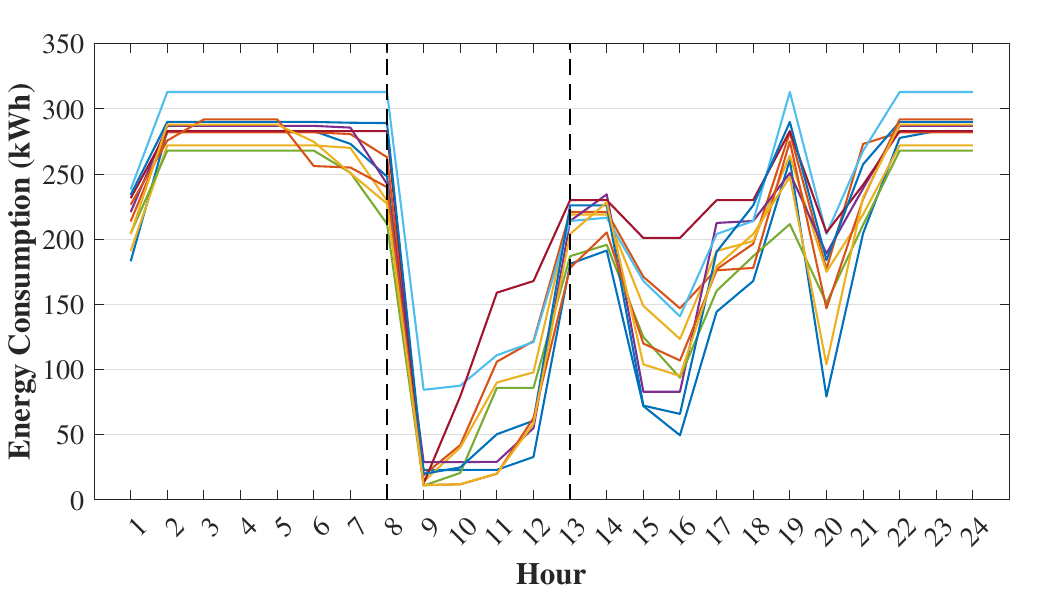}
    \includegraphics[width=1.5in]{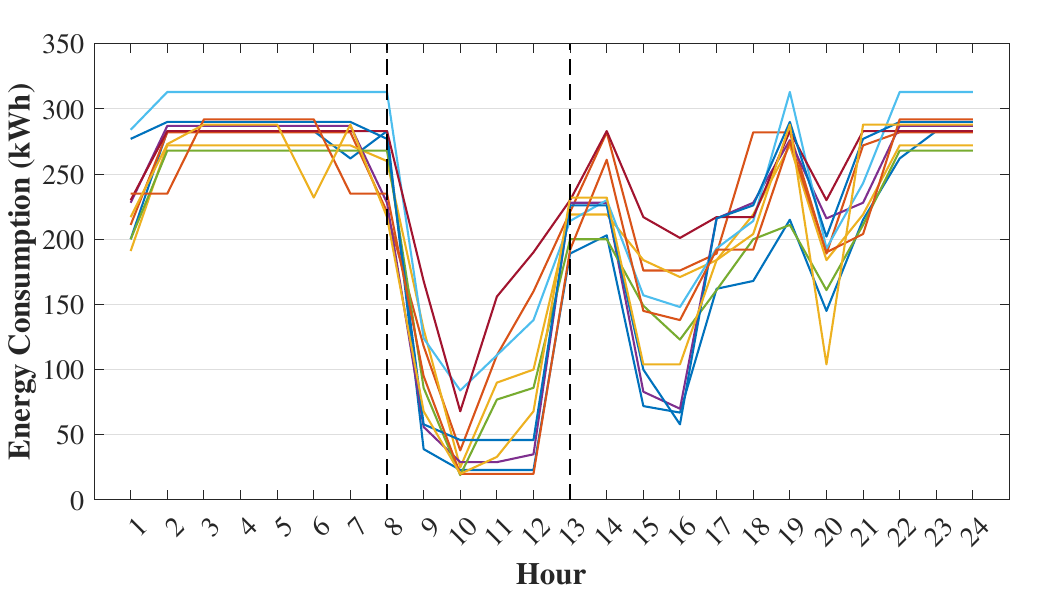}
\caption{Dispatch of the grid instructions among the factories: (left) LSTN. (right) STN.}
  \label{fig_dispatch}
\end{figure}

\begin{figure}[!t]
  \centering
  \includegraphics[width=2.5in]{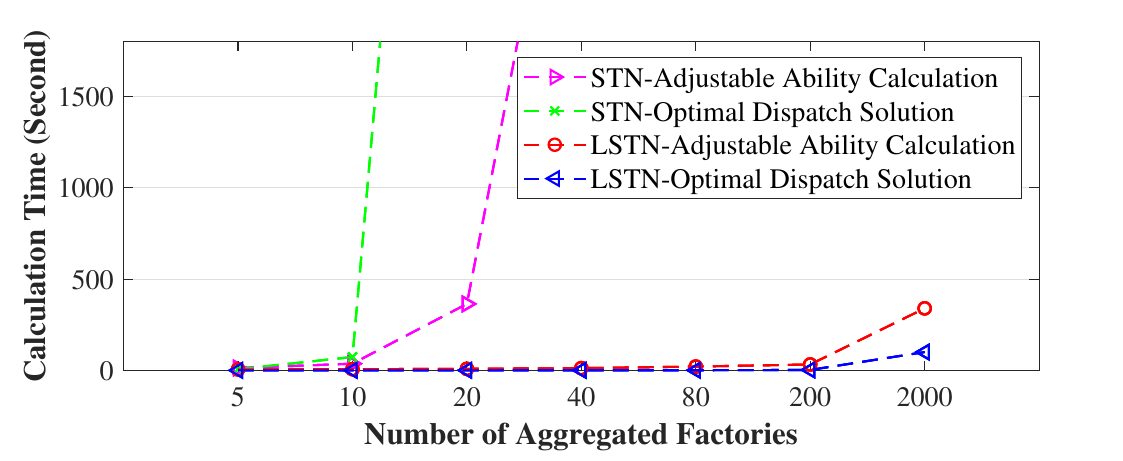}
\caption{Computation time with respect to the number of factories for calculating the adjustable ability and solving the optimal dispatch using the STN and LSTN.}
  \label{fig_calculation_time}
\end{figure}

Next, the grid sends instructions for reducing energy consumption according to the adjustable boundary of the aggregated factories. The dispatch instructions for the factories is shown in Fig.~\ref{fig_dispatch}. Fig.~\ref{fig_calculation_time} shows the computation time with respect to the number of factories for calculating the adjustable capacity and solving the optimal dispatch using STN-60 min and LSTN, respectively. STN-60 min failed to converge within 2 hours when the number of plants reached 20, with the number of binary variables exceeding 10000. In contrast, when the number of plants reaches 2000, the model with LSTN still resolved in minutes, representing an acceptable computation time for practical application and economic dispatch.

\section{Conclusion}\label{sec_conclusion}

Since the operational switching time of industrial equipment is usually much shorter than the scheduling time interval of DR, it is reasonable to assume equipment operation to be continuous in DR applications. Therefore, this paper presents a linear model to describe industrial production processes, denoted LSTN. Numerical results show that, compared to a standard STN model that accurately models the switching time of the operating state of the devices, the energy consumption error of LSTN is only 0.17\%. This error is much smaller than that of the conventional STN model with 1 hour as the scheduling time interval. Furthermore, because the linear model is more computationally tractable, LSTN can greatly accelerate the calculation of the aggregated DR capacity and the decomposition of grid instructions in the scenario of aggregating multiple factories, making it suitable for large-scale industrial DR.

\bibliographystyle{IEEEtran}
\bibliography{reference}

@article{zhuo_cost_2022,
	title = {Cost increase in the electricity supply to achieve carbon neutrality in {China}},
	volume = {13},
	copyright = {2022 The Author(s)},
	issn = {2041-1723},
	language = {en},
	number = {1},
	urldate = {2022-09-16},
	journal = {Nat Commun},
	author = {Zhuo, Zhenyu and Du, Ershun and Zhang, Ning and Nielsen, Chris P. and Lu, Xi and Xiao, Jinyu and Wu, Jiawei and Kang, Chongqing},
	month = jun,
	year = {2022},
	pages = {3172},
}

@article{li_precision_2022,
	title = {Precision and {Accuracy} {Co}-optimization {Based} {Demand} {Response} {Baseline} {Load} {Estimation} {Using} {Bi}-directional {Data}},
	issn = {1949-3061},
	doi = {10.1109/TSG.2022.3192386},
	journal = {IEEE Trans. Smart Grid},
	author = {Li, Kangping and Wang, Yuxi and Zhang, Ning and Wang, Fei},
	year = {2022},
	pages = {1--1},
}

@article{lu_data-driven_2021,
	title = {Data-driven real-time price-based demand response for industrial facilities energy management},
	volume = {283},
	issn = {0306-2619},
	language = {en},
	urldate = {2022-10-07},
	journal = {Appl. Energy},
	author = {Lu, Renzhi and Bai, Ruichang and Huang, Yuan and Li, Yuting and Jiang, Junhui and Ding, Yuemin},
	month = feb,
	year = {2021},
	pages = {116291},
}

@article{ding_demand_2014,
	title = {A {Demand} {Response} {Energy} {Management} {Scheme} for {Industrial} {Facilities} in {Smart} {Grid}},
	volume = {10},
	issn = {1941-0050},
	doi = {10.1109/TII.2014.2330995},
	number = {4},
	journal = {IEEE Trans. Ind. Informat.},
	author = {Ding, Yue Min and Hong, Seung Ho and Li, Xiao Hui},
	month = nov,
	year = {2014},
	pages = {2257--2269},
}

@article{kondili_general_1993,
	title = {A general algorithm for short-term scheduling of batch operations—{I}. {MILP} formulation},
	volume = {17},
	issn = {00981354},
	language = {en},
	number = {2},
	urldate = {2022-10-09},
	journal = {Computers \& Chemical Engineering},
	author = {Kondili, E. and Pantelides, C.C. and Sargent, R.W.H.},
	month = feb,
	year = {1993},
	pages = {211--227},
}

@inproceedings{zhang_industrial_2015,
	title = {Industrial demand response by steel plants with spinning reserve provision},
	doi = {10.1109/NAPS.2015.7335115},
	booktitle = {2015 {North} {American} {Power} {Symposium} ({NAPS})},
	author = {Zhang, Xiao and Hug, Gabriela and Kolter, Zico and Harjunkoski, Iiro},
	month = oct,
	year = {2015},
	pages = {1--6},
}

@article{yu_real-time_2016,
	title = {A real-time decision model for industrial load management in a smart grid},
	volume = {183},
	issn = {0306-2619},
	language = {en},
	urldate = {2022-11-09},
	journal = {Appl. Energy},
	author = {Yu, Mengmeng and Lu, Renzhi and Hong, Seung Ho},
	month = dec,
	year = {2016},
	pages = {1488--1497},
}

@article{cortez_demand_2023,
	title = {Demand {Management} for {Peak} to {Average} {Ratio} {Minimization} via {Intraday} {Block} {Pricing}},
	issn = {1949-3053, 1949-3061},
	language = {en},
	urldate = {2023-02-09},
	journal = {IEEE Trans. Smart Grid},
	author = {Cortez, Carolina and Kasis, Andreas and Papadaskalopoulos, Dimitrios and Timotheou, Stelios},
	year = {2023},
	pages = {1--1},
}

@article{tan_optimal_2023,
	title = {Optimal {Virtual} {Battery} {Model} for {Aggregating} {Storage}-{Like} {Resources} with {Network} {Constraints}},
	language = {en},
	journal = {CESS},
	author = {Tan, Zhenfei and Yu, Ao and Zhong, Haiwang and Zhang, Xianfeng and Xia, Qing and Kang, Chongqing},
	month = mar,
	year = {2023},
}

@article{castro_resourcetask_2013,
	title = {Resource–{Task} {Network} {Formulations} for {Industrial} {Demand} {Side} {Management} of a {Steel} {Plant}},
	volume = {52},
	issn = {0888-5885, 1520-5045},
	language = {en},
	number = {36},
	urldate = {2023-03-14},
	journal = {Ind. Eng. Chem. Res.},
	author = {Castro, Pedro M. and Sun, Lige and Harjunkoski, Iiro},
	month = sep,
	year = {2013},
	pages = {13046--13058},
}

@article{nolde_electrical_2010,
	title = {Electrical load tracking scheduling of a steel plant},
	volume = {34},
	issn = {00981354},
	language = {en},
	number = {11},
	urldate = {2023-03-14},
	journal = {Computers \& Chemical Engineering},
	author = {Nolde, Kristian and Morari, Manfred},
	month = nov,
	year = {2010},
	pages = {1899--1903},
}

@article{floudas_continuous-time_2004,
	title = {Continuous-time versus discrete-time approaches for scheduling of chemical processes: a review},
	volume = {28},
	issn = {00981354},
	shorttitle = {Continuous-time versus discrete-time approaches for scheduling of chemical processes},
	language = {en},
	number = {11},
	urldate = {2023-03-14},
	journal = {Computers \& Chemical Engineering},
	author = {Floudas, Christodoulos A. and Lin, Xiaoxia},
	month = oct,
	year = {2004},
	pages = {2109--2129},
}

@article{samad_smart_2012,
	series = {{FOCAPO} 2012},
	title = {Smart grid technologies and applications for the industrial sector},
	volume = {47},
	issn = {0098-1354},
	language = {en},
	urldate = {2023-03-24},
	journal = {Computers \& Chemical Engineering},
	author = {Samad, Tariq and Kiliccote, Sila},
	month = dec,
	year = {2012},
	pages = {76--84},
}

@article{wen_frequency_2016,
	title = {Frequency {Dynamics} {Constrained} {Unit} {Commitment} {With} {Battery} {Energy} {Storage}},
	volume = {31},
	issn = {1558-0679},
	doi = {10.1109/TPWRS.2016.2521882},
	number = {6},
	journal = {IEEE Trans. Power Syst.},
	author = {Wen, Yunfeng and Li, Wenyuan and Huang, Gang and Liu, Xuan},
	month = nov,
	year = {2016},
	pages = {5115--5125},
}

@inproceedings{piphitpattanaprapt_optimal_2015,
	title = {Optimal dispatch strategy of hybrid power generation with battery energy storage system in islanding mode},
	doi = {10.1109/ISGT-Asia.2015.7387153},
	booktitle = {2015 {IEEE} {Innovative} {Smart} {Grid} {Technologies} - {Asia} ({ISGT} {ASIA})},
	author = {Piphitpattanaprapt, Noppasit and Bangerdpongchai, David},
	month = nov,
	year = {2015},
	pages = {1--6},
}

@article{lu_multi-agent_2020,
	title = {Multi-agent deep reinforcement learning based demand response for discrete manufacturing systems energy management},
	volume = {276},
	issn = {0306-2619},
	language = {en},
	urldate = {2022-10-31},
	journal = {Appl. Energy},
	author = {Lu, Renzhi and Li, Yi-Chang and Li, Yuting and Jiang, Junhui and Ding, Yuemin},
	month = oct,
	year = {2020},
	pages = {115473},
}

@article{wohlfarth_demand_2020,
	title = {Demand response in the service sector – {Theoretical}, technical and practical potentials},
	volume = {258},
	issn = {03062619},
	language = {en},
	urldate = {2022-12-25},
	journal = {Appl. Energy},
	author = {Wohlfarth, Katharina and Klobasa, Marian and Gutknecht, Ralph},
	month = jan,
	year = {2020},
	pages = {114089},
}

@article{zhang_demand_2018,
	title = {Demand {Response} of {Ancillary} {Service} {From} {Industrial} {Loads} {Coordinated} {With} {Energy} {Storage}},
	volume = {33},
	language = {en},
	number = {1},
	urldate = {2022-10-30},
	journal = {IEEE Trans. Power Syst.},
	author = {Zhang, Xiao and Hug, Gabriela and Kolter, J. Zico and Harjunkoski, Iiro},
	month = jan,
	year = {2018},
	pages = {951--961},
}

@inproceedings{Lofberg2004,
address = {Taipei, Taiwan},
author = {L{\"{o}}fberg, J.},
booktitle = {In Proc. of the CACSD Conf.},
title = {YALMIP : A Toolbox for Modeling and Optimization in MATLAB},
year = {2004}
}

\end{document}